\documentclass[aps,prl,reprint, superscriptaddress]{revtex4-2}
\usepackage{blindtext}
\usepackage{graphicx}
\usepackage{xcolor}
\usepackage{dcolumn}
\usepackage{bm}
\usepackage[utf8]{inputenc}
\usepackage{changes}
\usepackage{natbib}

\begin{document}
	
	\preprint{APS/123-QED}
	
	\title{Multi band Fermi surface in 1T-VSe$_{2}$ and its implication for charge density wave phase}

\author{Turgut Yilmaz}
\affiliation{National Synchrotron Light Source II, Brookhaven National Lab, Upton, New York 11973, USA}
\email{tyilmaz@bnl.gov}

\author{Boris Sinkovic}
\affiliation{Department of Physics, University of Connecticut, Storrs, Connecticut 06269, USA}

\author{Elio Vescovo}
\affiliation{National Synchrotron Light Source II, Brookhaven National Lab, Upton, New York 11973, USA}

\date{\today}

\begin{abstract}
	
	Here, our angle resolved photoemission spectroscopy experiment reveled that the surface band structure of the 1T-VSe$_2$ host electronic states that was not predicted or probed before. Earlier claims to support charge density wave phase can be all explained in terms of these new findings. Its Fermi surface found to be not gaped at any point of the Brillouin zone and warping effect on the electronic structure, attributed to the lattice distortion previously, is due to the different dispersion of the multiple bands. Based on these new findings and interpretations, charge density wave induced modification on the electronic structure of 1T-VSe$_2$ needs to be reconstructed in the future studies.

\end{abstract}

\maketitle

Layered transition metal dichalcogenides (TMDCs) host various electronic, structural, and transport phenomena making them most promising candidates for applications in electronic devices \cite{mak2016photonics, manzeli20172d}. Among the many novel states, the charge density wave (CDW) phase in these compounds is particularly under scrutiny as it is neighbor to superconductivity in the phase diagrams \cite{cho2018using, yu2021unusual}. In terms of electronic structure, the CDW phase is usually linked to Fermi surface nesting with opening of a gap at specific portion of the Fermi surface connected by the CDW wave vector \cite{gruner1988dynamics, monceau2012electronic, wilson1974charge}. In response, atoms move from their original positions forming a superstructure that can be visualized through a scanning tunneling electron microscopy (STM) experiment \cite{wang1991spectroscopy, giambattista1990scanning, eaglesham1986charge}.

Of all TMDCs, 1T-VSe$_2$ is a special example due to its long wavelength 3D CDW phase in the bulk. It undergoes an incommensurate CDW with $4a\times4a\times3.18c$ periodic lattice distortion around $T^*$ = 110 K, followed by a second transition to a commensurate CDW state around 80 K \cite{giambattista1990scanning, eaglesham1986charge, tsutsumi1982x}. On the other hand, CDW-phase on the electronic structure is anomalously supported by angle resolved photoemission spectroscopy (ARPES) studies. Such as, CDW induced gap has been reported around the $\overline{M}$-point of the Fermi surface as a shift of a secondary peak while a spectral intensity suppression at the Fermi level is absent \cite{terashima2003charge, sato2004three, wang2021three, falke2021coupling}. This appeared to be a well established way of measuring the CDW induced gap in the bulk 1T-VSe$_2$ samples. Furthermore, a recent STM study also reported a gap at the Fermi surface which is found to be more pronounced around the $\Gamma$-point \cite{jolie2019charge}. However, this observation should not be correlated with the structural transition since the $\Gamma$-point is not along the CDW wave vector. The results are even more elusive in the monolayer VSe$_2$ samples. First, fully gaped Fermi surface claimed to be reported by multiple studies \cite{chen2018unique, feng2018electronic, zong2022observation}. Such observation is unlikely to be due to the CDW phase because CDW wave vector clearly has no influence in that case. Second, monolayer samples seem to have a larger CDW induced gap ($\sim$100 meV) with a higher T$^*$. On the other hand, experimental band structures do not exhibit any signature of back-bending to support such a large energy gap in contrast to the computational models \cite{chen2018unique, feng2018electronic, zong2022observation}.

Here, we re-evaluate some of the main results reported on 1T-VSe$_2$ by performing a detailed ARPES experiment and rule out the earlier CDW related claims point by point. The Fermi surface of 1T-VSe$_2$ contains multiple bands, not previously detected and they have different in-plane and out-plane dispersion. Thereby, the previous observations that are attributed to the CDW phase are due to the lack of resolving these bands. In this regard, the previous theoretical models and experimental observations cannot be assigned to the CDW phase and the impact of structural distortion on the electronic structure requires further characterizations and approaches. But, new observations provide more detailed electronic structure that can guide future studies as well as suggesting the revision of the various other studies regarding electronic properties of 1T-VSe$_2$.

\begin{figure*}[t]
	\centering
	\includegraphics[width=16cm,height=5.283cm]{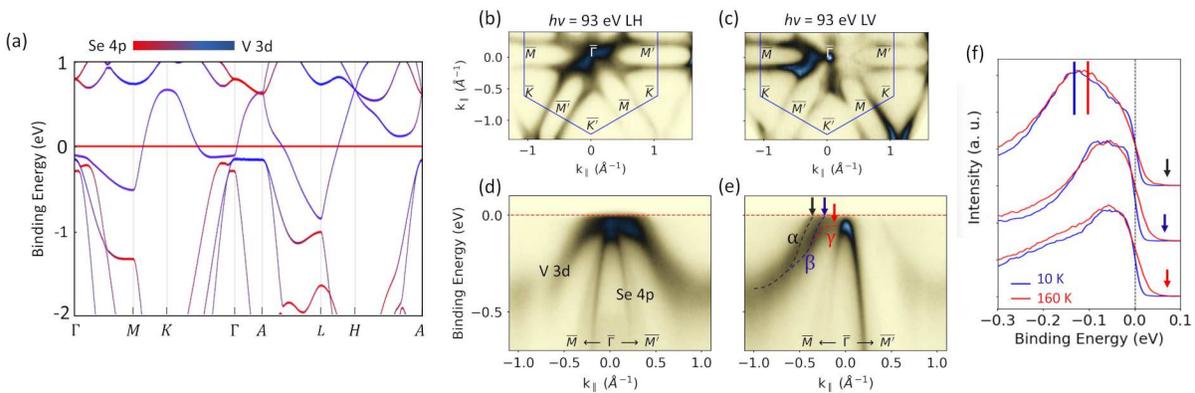}
	\caption{
		(a) Orbital projected band structure of the bulk VSe$_2$ in normal state. Red horizontal line represents the Fermi level. (a-c) Fermi surfaces of a bulk 1T-VSe$_2$ sample were taken with LH at 10 K, LV at 10, and LV at 160 K, respectively. (d-f) Corresponding ARPES maps along the $\overline{M}$- $\overline{\Gamma}$ - $\overline{M'}$ direction. The asymmetric band structure observed with LV polarized lights is likely due to the 3-fold rotational symmetry of the band structure. The orbital character of the bands is adapted from the literature (g) EDCs taken along the black, blue, and red arrows shown in (e).
	}
\end{figure*}

Single crystal 1T-VSe$_2$ were obtained from 2dsemiconductors company. ARPES experiments were performed at 21-ID-1 (ESM) beamline of NSLS-II using a DA30 Scienta electron spectrometer with an energy resolution better than 12 meV and the beam spot size was approximately 5 $\mu$m$^2$. The synchrotron radiation incidence angle was 55$^o$. Analyzer slit was along to the $\overline{M}$ - $\overline{\Gamma}$ - $\overline{M'}$ direction during the ARPES measurements at normal emission and along to the  $\overline{M}$ - $\overline{M'}$-direction during the zone corner scans. LV polarized light is parallel to the sample surface and analyzer slit while LH polarized light is on the incident plane. Band structure calculation is performed with Quantum Espresso package \cite{giannozzi2009quantum, giannozzi2017advanced}. The photon energy is converted to k$_z$ crystal momentum adopting the free electron final state approximation given as $\hbar$k$_z$ = 2m$_e$$\sqrt{E_{kin}\cos(\theta)+V_o}$ 
where m$_e$ is the free electron mass, E$_{kin}$ is the photoelectron kinetic energy, and V$_o$, inner potential, set to 7.5 eV \cite{strocov2012three}.

We first investigate computed electronic structure in Figure 1(a) to compare it with the experimental one. Fermi level is dominated by V 3$d$ atomic orbitals which forms a V-saped bulk bands at $M(L)$-points while it exhibits relatively flatter feature near the zone center. Highly dispersive Se 4$p$ atomic orbitals mostly located around the $\Gamma$($A$)-points below the Fermi level. Furthermore the V 3$d$ atomic orbitals do not exhibit any splitting near the $\Gamma$($A$) or $M(L)$-points. Particularly, a single band is crossing the Fermi level in the vicinity of the $M(L)$-point. This band structure is in excellent agreement with the previous studies \cite{he2021confinement, coelho2019charge}, but it underestimates the details as we will discuss latter in this work.

The experimental electronic structure of 1T-VSe$_2$ bulk samples is studied by using linear horizontal (LH) and linear vertical (LV) polarized lights in Figure 1(b-e), respectively. In the LH geometry, apparent ellipsoidal electron pockets centered at $\overline{M}$($\overline{M'}$)-points are the main features of the Fermi surface while highly dispersive Se 4$p$ atomic orbitals dominate the zone center (Figure 1(b)). The corresponding ARPES map along the $\overline{M}$-$\overline{\Gamma}$-$\overline{M'}$ exhibits V-3$d$ atomic orbitals with nearly flat dispersion in the vicinity of the $\overline{\Gamma}$-point where it overlaps with Se 4$p$ atomic orbitals (Figure 1d). Overall band structure obtained for this particular experimental geometry is consistent with the computed one as well as the earlier ARPES reports \cite{wang2021three, strocov2012three, chen2018unique, chen2018unique, coelho2019charge}.

On the other hand, the same band structure taken with LV polarized light reveals differences in details. The zone center exhibits a point like dispersion enabling to study the band structure around the $\overline{\Gamma}$-point more precisely. Unlike the single band observed with LH polarized light, three bands called $\alpha$, $\beta$, and $\gamma$ are now resolved  around the zone center (Figure 1(e)). The first two bands are crossing the Fermi level at k$_{\parallel}$ = -0.37 $\AA$$^{-1}$ and at k$_{\parallel}$ = -0.24 $\AA$$^{-1}$ while the third band is located just below the Fermi level displaying a flatter dispersion towards to $\overline{\Gamma}$-point. These additional states are not predicted in the band structure calculation reported here or elsewhere \cite{wang2021three, strocov2012three, chen2018unique, chen2018unique, coelho2019charge, he2021confinement, coelho2019charge}. Implication of the new findings in terms of CDW induced gap is that if a band is located just below the Fermi level, temperature induced lifetime broadening can always lead to a determination of a false gap. This is likely to be the case for a recent STM study that reports a prominent gap in the vicinity of the $\overline{\Gamma}$-point \cite{jolie2019charge}. Furthermore, these multiple bands are main characteristics of the material and can be resolved below and above the T$^*$ showing that they are not due to the structural distortion(Figure 1(c) and Supplementary Materials Figure S1(a-b)).

By using this new finding, one can reproduce a data appear to host a temperature dependent gap at the Fermi level. The EDCs crossing the $\alpha$ and $\beta$ bands at the Fermi level and momentum point across the $\gamma$ band are given in Figure 1(f) for two temperatures; 10 K (below T$^*$ and 160 K (above T$^*$). Third one has a natural gap since the band is located below the Fermi level and the second one clearly shows no evidence of a CDW gap. However, first one appears to experiencing a shift towards higher binding energy at a lower temperature (marked with blue and red vertical lines in Figure 1(f)). Such shift commonly attributed to the a gap opening in this material \cite{terashima2003charge, sato2004three, falke2021coupling}. But, spectral intensity at the Fermi level does not show prominent temperature dependence. This apparent shift is due to the $\beta$ band being located just below the $\alpha$ band. And, temperature induced lifetime broadening leads them to overlap more in energy. This appears in the band structure as if a band is shifting in energy with temperature. Thereby, lack of the resolving these bands in the earlier studies leads to misinterpreting the experimental data. Indeed, this should have been realized since the vicinity of the $\Gamma$-point is not along the CDW-wave vector and hence is not expected to be gaped.

Furthermore, the modulation of the color contrast on the electronic structure figures is relied on as a signature of a gap opening at the Fermi level \cite{sato2004three, wang2021three}. Here, we repeat a similar experiment. Figure 2(a) presents the Fermi surface that covers the ellipsoidal electron pocket centered at the $\overline{M}$-point. Color-contrast is gradually decreasing towards $\overline{M}$-point as consistent with the earlier studies. In this regard, one can claim a possible gap opening at the Brillouin zone boundaries. To check if this is the case, a logical experimental approach will be comparing the Fermi edges taken from the different portions of the Fermi surface recorded below T$^*$. In this way, possible experimental errors induced by variable temperature can be also avoided. Thus, ARPES maps taken along the dashed black line in Figure 2(a) and the $\overline{K}$ - $\overline{M}$ - $\overline{K'}$ direction are given in Figure 2(b-c), respectively. Compared to the former one, the latter exhibits lower spectral intensity in the entire band structure rather than only at the Fermi level. Therefore, the apparent spectral intensity does not necessarily represent a gap opening.

\begin{figure}[t!]
	\centering
	\includegraphics[width=8.5cm,height=6.502cm]{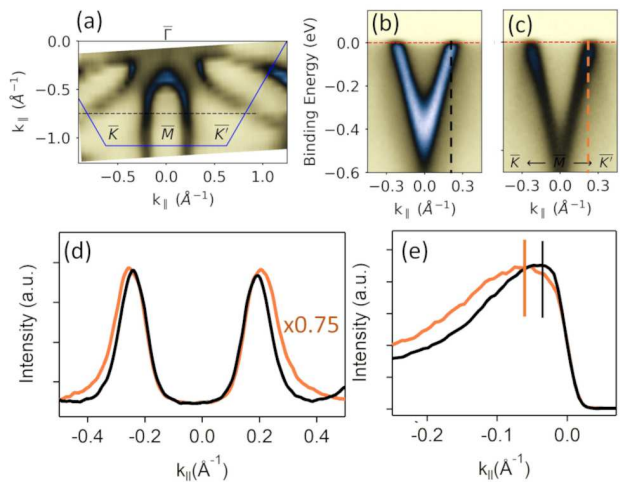}
	\caption{
		(a) The Fermi surface is recorded with 93 eV LH polarized lights. (b-c) ARPES maps taken along the dashed black line in (a) and $\overline{K}$- $\overline{M}$ - $\overline{K'}$, respectively. (d) MDCs were taken at the Fermi level from (b-c). (e) EDCs taken along the dashed black and orange lines in (b-c). Vertical black and orange lines in (e) mark the high intensity points. (f-g) ARPES maps along the  $\overline{M}$ - $\overline{M'}$ direction taken
		with 101 eV LH and LV polarized lights, respectively. All data are collected at 10 K.}
\end{figure}

The MDCs taken at the Fermi level of Figure 2(b-c) are compared in Figure 2(d) to find out the origin of this spectral intensity modulation. The one along the $\overline{K}$ - $\overline{M}$ - $\overline{K'}$ direction exhibits a broader line shape compared to the other one. Therefore, the lower intensity around $\overline{M}$ of the Fermi surface is not due to the gap opening but it is caused by a broader electronic structure in momentum space. Furthermore, the EDCs taken along the dashed black line in Figure 2(b) and the dashed orange line in Figure 2(c) are given in Figure 2(e). They exhibit the same Fermi level indicating the absence of a insulating gap. On the other hand, the high intensity points show a difference in the EDCs (vertical orange and black lines in Figure 2(e)). This apparent shift has been considered as well-established experimental signature of the CDW induced gap in VSe$_2$ \cite{terashima2003charge, sato2004three, wang2021three, falke2021coupling}. Indeed, these observations indicate existence of multiple bands located in close proximity to each other rather than a gap opening. Therefore, the Fermi surface is expected to be dominated at least two bands in contrast to the previous experiments and computations.

Above argument is experimentally confirmed by analysing the various ARPES maps parallel to $\overline{K}$ - $\overline{M}$ - $\overline{K}'$. Around $\overline{M}$-point a V-shape band is crossing the Fermi level at k$_\vert$ = $\pm$ 0.32 $\AA$ (Figure 3(b)). However, band splitting towards zone center can be seen in Figure 3(c-e) as marked with blue arrows. These additional bands eventually connect with the neighbor electron pockets (Figure 3(f-g)). 

\begin{figure}[t!]
	\centering
	\includegraphics[width=8.5cm,height=6.418cm]{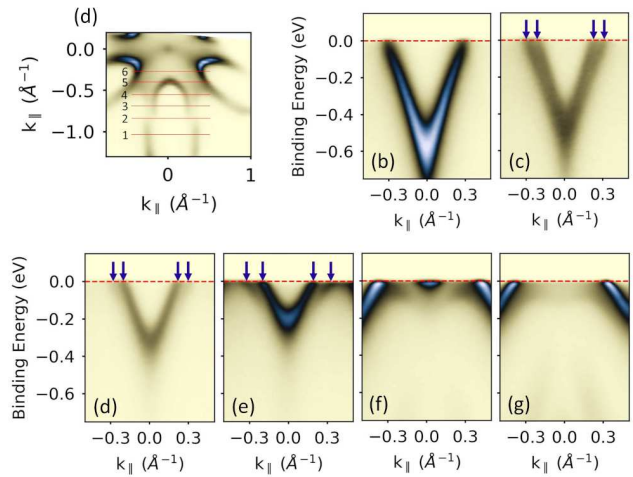}
	\caption{
		(a) The Fermi surface. (b-g) ARPES cuts along the 1-5 marked lines shown in (a), respectively. Spectrum in (b) is along the $\overline{K}$- $\overline{M}$ - $\overline{K'}$ direction and (d-g) is taken parallel to it. Blue arrows show the double bands. All spectra collected with 75 eV LH polarized light at 10 K.}
\end{figure}

\begin{figure}[t!]
	\centering
	\includegraphics[width=8.5cm,height=10.312cm]{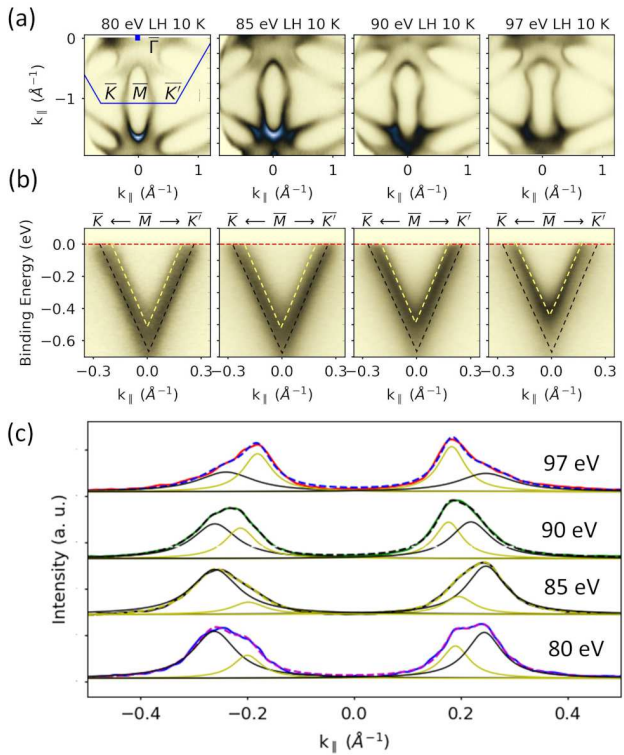}
	\caption{
		(a-b) The Fermi surfaces and corresponding ARPES maps along $\overline{K}$- $\overline{M}$ - $\overline{K'}$ direction. (c) MDCs taken at the Fermi level from each spectrum. Orange and black lines in (c) are the Lorentzian peak fittings and dashed lines in (c) are sum of fittings. }
\end{figure}

Multiple bands can be also probed in the band structure along the $\overline{K}$ - $\overline{M}$ - $\overline{K}'$ with more elaborate approach. Fermi surfaces taken at various photon energies are given in Figure 4(a). In the data, 80 eV and 97 eV photons correspond to k$_z$ = $A$ and k$_z$ = $\Gamma$ high symmetry points in 2D momentum plane. Ellipsoidal pocket centered at $\overline{M}$-point shows a strong photon energy dependence and it resembles more like a dog-bone at k$_z$ = $\Gamma$. In addition to this strong spectral feature, second band with a weaker intensity can be also visually observed in some parts of the Fermi surface. These multiple bands can be seen in ARPES maps along $\overline{K}$ - $\overline{M}$ - $\overline{K}'$ as marked with dashed yellow and black lines (Figure 4(b)).

To further support this observation, the MDCs taken at the Fermi level are compared in Figure 4(c). Bands located at $\pm$ k$_\vert$ with respect to  $\overline{M}$-point exhibit double peak features as indicated with yellow and black fitting lines. Thereby, the V-shape conduction band is indeed formed by two bands as inner and outer ones. Based on Figure 3(b-c), these two bands have distinct k$_z$ dispersion as the outer band is 2D like while the inner one has 3D character and their spectral intensities show an opposite k$_z$ dependence. This has an important implication for the CDW phase of the VSe$_2$. One of the fundamental experimental evidence for the CDW phase was reported to be a warping effect on the Fermi surface along the k$_z$ direction \cite{sato2004three, strocov2012three, wang2021three}. However, this warping effect is clearly due to the different nature of these mulitple bands crossing the Fermi level and not correlated with any type of structural distortion. Indeed, the appearent warping of the band structure is temperature independent (Supplementary Materials SFigure 2). Therefore, earlier observations are due to the lack of resolving multiple bands leading to misinterpret the experimental data.

In summary, unlike the earlier studies, we show that the Fermi surface of VSe$_2$ is dominated by multiple bands. By using this, each of the previous observation can be explained and shown to be not related to any structural distortion. These new findings need to be considered in order to understand the correct nature of the CDW phase and find the structural distortion induced modification on the electronic structure. Furthermore, these realization also suggest that the large CDW induced gap of 100 meV  in the monolayer VSe$_2$ is questionable \cite{chen2018unique, feng2018electronic, zong2022observation, chen2018unique, feng2018electronic, zong2022observation}. This is discussed in the Supplementary Materials SFigure 3 and no gap is deducted around the $\overline{M}$-point that can be associated with CDW phase. Furthermore, the apparent width of the Fermi edge around the $\overline{M}$-point does not change with the temperature in the monolayer case. This indicates that it is dominated by the temperature independent defects such as vacancies which leads misinterpretation of the data. All these new findings needs to be considered in the future studies and they are probably the main reason for why the mechanism behind the 3D charge density wave phase in VSe$_2$ has never been understood.

This research used resources ESM (21ID-I) beamline of the National Synchrotron Light Source II, a U.S. Department of Energy (DOE) Office of Science User Facility operated for the DOE Office of Science by Brookhaven National Laboratory under Contract No. DE-SC0012704. We have no conflict of interest, financial or other to declare.

\bibliographystyle{plain}

\end{document}